\documentclass[aps,prl,twocolumn,showpacs,groupedaddress]{revtex4}

\usepackage{epsfig}
\usepackage{amssymb}

\newcommand{\om}{\omega}

\newcommand{\beq}{\begin{equation}}
\newcommand{\eeq}{\end{equation}}
\newcommand{\ba}{\begin{array}}
\newcommand{\bea}{\begin{eqnarray}}
\newcommand{\ea}{\end{array}}
\newcommand{\eea}{\end{eqnarray}}

\newcommand{\pa}{\parallel}
\newcommand{\pe}{\perp}

\newcommand\comment[1]{ \hbox{[{\it Comment suppressed here.}\/]} }
\newcommand\hide[1]{}

\newcommand{\skipover}[1]{}

\newcommand{\calO}{{\cal O}}

\newcommand{\Tr}{\hbox{Tr}}

\newcommand{\bp}{{\bf p}}
\newcommand{\bq}{{\bf q}}

\newcommand{\be}{\begin{equation}}
\newcommand{\ee}{\end{equation}}
\newcommand{\bean}{\begin{eqnarray*}}
\newcommand{\eean}{\end{eqnarray*}}

\newcommand{\C}{\mathcal{C}}

\begin{document}

\title{Parametric resonance in quantum field theory}

\author{J{\"u}rgen Berges}
\author{Julien Serreau}
\affiliation{Institute for Theoretical Physics, Heidelberg University\\
Philosophenweg 16, 69120 Heidelberg, Germany}

\begin{abstract}
\noindent
We present the first study of parametric resonance in quantum field theory 
from a complete next-to-leading order calculation in a $1/N$-expansion
of the 2PI effective action, which includes scattering and memory effects.
We present a complete numerical solution for an $O(N)$-symmetric scalar 
theory and provide an approximate analytic description of the nonlinear 
dynamics in the entire amplification range.  
We find that the classical resonant amplification at early times
is followed by a collective amplification regime with
explosive particle production in a broad momentum range, which
is not accessible in a leading-order calculation.
\end{abstract}
\pacs{11.15.Pg, 98.80.Cq, 05.70.Ln \hfill HD--THEP--02-44}

\maketitle

\noindent
In quantum field theory the phenomenon of parametric resonance describes 
the resonant amplification of quantum fluctuations, 
which can be interpreted as particle 
production. It provides an important building block for our understanding 
of the (pre)heating of the early universe after a period of 
inflation~\cite{preheat1}. It has been frequently discussed  
for relativistic heavy-ion collisions in 
the formation of disoriented chiral condensates~\cite{dcc}, or the 
decay of Polyakov-loop condensates~\cite{Ploop}, or parity-odd 
bubbles~\cite{bubbles}.

Despite being a basic phenomenon that can occur in a large variety
of quantum field theories, parametric resonance is a rather complex 
process, which has so far defied most attempts for a complete 
analytic treatment
even for simple theories. It is a far-from-equilibrium phenomenon 
involving densities inversely proportional to the coupling. 
The nonperturbatively large occupation numbers cannot be described by 
standard kinetic descriptions. So far, classical statistical 
field theory simulations on the lattice have been the only quantitative 
approach available~\cite{preheat2}. These are valid for 
not too late times, before the approach to quantum thermal 
equilibrium sets in. Up to now, studies in quantum field theory have been 
mainly limited to linear or mean-field type approximations (leading-order 
in large-$N$, Hartree)~\cite{LOapp}, which present a valid description
for sufficiently early times. However, they are known 
to fail to describe thermalization and miss important rescattering 
effects~\cite{preheat2,preheat3}.   
Going beyond mean-field has long been a major difficulty in practice: 
Similar to perturbation theory, standard approximations such as based on 
$1/N$--expansions of the one-particle irreducible (1PI) effective action 
can be secular in time and do not provide a valid description. 
In contrast, it has recently been demonstrated~\cite{quantum,thermal} 
for $1+1$ dimensional theories that 
far-from-equilibrium dynamics and subsequent thermalization can
be described using a $1/N$-expansion
of the {\em two-particle irreducible} (2PI) effective 
action~\cite{2PI,quantum,AABBS}. Below, we show that
this provides a systematic and {\em practicable} 
nonperturbative approach for quantum field
theories in $3+1$ dimensions and with a nonzero macroscopic 
field~$\phi$, relevant for realistic 
particle physics applications.

In this work, we present the first quantum field theoretical 
description of the phenomenon of parametric resonance 
taking into account rescattering: For an $O(N)$--symmetric scalar 
field theory we employ the 2PI $1/N$-expansion to next-to-leading order 
(NLO), which includes off-shell and memory effects~\cite{quantum,AABBS}. 
We point out that the approach solves the problem of an explicit description 
of the dynamics of correlation functions at nonperturbatively large densities. 
We present a complete numerical solution of the corresponding equations of 
motion. Moreover, we identify the relevant contributions to the dynamics at 
various times and provide an approximate analytic description of the 
nonlinear dynamics for the entire amplification range.
Apart from the resonant amplification in the linear regime, we identify 
two characteristic time scales, which signal strongly enhanced particle 
production in a broad momentum range due to nonlinear, source effects.  
This collective amplification is crucial for the rapid approach to a 
subsequent, quasistationary regime, where direct scattering drive a very 
slow evolution towards thermal equilibrium. We emphasize that these processes 
cannot be seen in mean-field approximations. Similar phenomena have been 
observed in classical field theories~\cite{preheat2}, and we present an 
analytic criterion for the validity of classical statistical approximations.
These effects are important for a reliable 
description of the system at the end of the resonance stage for finite 
$N \lesssim 1/\lambda$. For realistic inflationary models with typically 
$\lambda \ll 1$ this is, in particular, crucial to determine whether there 
are any radiatively restored symmetries.

\begin{figure}
 \centerline{\epsfxsize=8.cm\epsffile{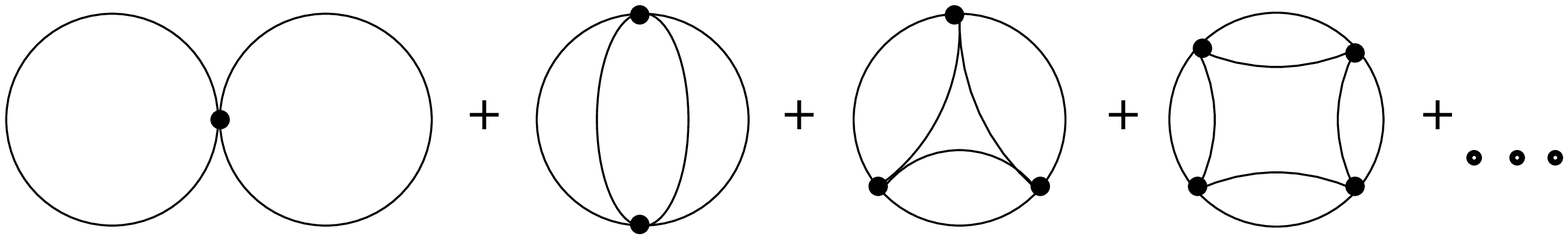}}
 \vspace{.3cm}
 \centerline{\epsfxsize=6.5cm\epsffile{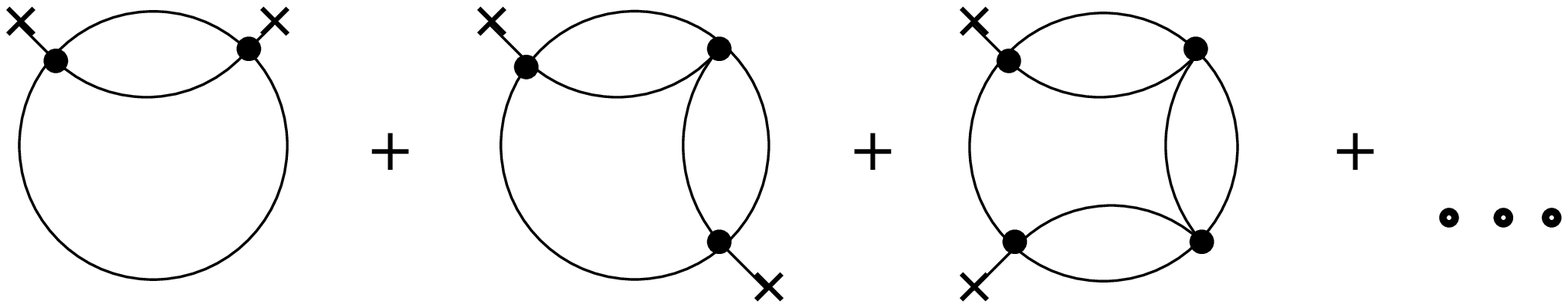}}
 \caption{\label{fig:SSBNLO} 
 The dots indicate that each diagram is obtained from the previous 
 one by adding another ``rung'' with two full propagator lines at each vertex.
 The crosses denote field insertions.\vspace*{-0.4cm}}
\end{figure}

We consider a relativistic real scalar field 
$\varphi_a$ ($a\!=\!1,\ldots, N$)
with action 
$S[\varphi]=- \int_x \{ \frac{1}{2}\varphi_a(\square_x + m^2)\varphi_a+
\frac{\lambda}{4! N}\, (\varphi_a\varphi_a)^2 \}$, 
where summation over repeated indices is implied.
We use the notation $\int_x \equiv \int_\C {\rm d}x^0 \int {\rm d}{\bf x}$
with $\C$ denoting a closed time path along the real axis.
All correlation functions of the quantum theory can be obtained from the 
2PI generating functional for Green's functions $\Gamma[\phi,G]$, parametrized 
by the field expectation value 
$\phi_a(x)=\langle\varphi_a(x)\rangle$ 
and the connected propagator 
$G_{ab}(x,y)=\langle {\rm T}_{\C} \varphi_a(x)\varphi_b(y)\rangle 
- \phi_a(x) \phi_b(y)$ 
\cite{2PI}:
$$
\Gamma[\phi,G] = S[\phi] + \frac{i}{2} \Tr\ln G^{-1} 
          + \frac{i}{2} \Tr\, G_0^{-1}(\phi)\, G
          + \Gamma_2[\phi,G] \, ,
$$ 
where $ i G^{-1}_{0,ab}(x,y;\phi) \equiv 
\frac{\delta^2 S[\phi]}{\delta\phi_a(x)\delta\phi_b(y)}$.
The term $\Gamma_2[\phi,G]$ contains all 
contributions beyond one-loop order and can be represented as 
a sum over closed 2PI graphs~\cite{2PI}. To NLO 
in the 2PI $1/N$-expansion, $\Gamma_2[\phi,G]$ contains the diagrams 
with topology shown in Fig.~\ref{fig:SSBNLO}~\cite{quantum,AABBS}. 
The infinite series can be summed analytically:
\bea 
\label{NLOcont} 
\lefteqn{
\Gamma_2[\phi,G] =  -\frac{\lambda}{4!N}
\int_{x} G_{aa}(x,x)G_{bb}(x,x) + \frac{i}{2}\,\Tr\,  
\mbox{ln} \, {\bf B}(G) }  \nonumber \\ 
&+& \!\!\frac{i\lambda^2}{(6N)^2} \int_{xyz} \! 
{\bf B}^{-1}(x,z;G) G^2(z,y) 
\phi_a(x) G_{ab}(x,y) \phi_b (y) \nonumber
\eea
with ${\bf B}(x,y;G) \equiv \delta (x-y)
 + \frac{i \lambda}{6 N} G^2(x,y)$ and 
$G^2(x,y) \equiv G_{ab}(x,y)G_{ab}(x,y)$. 
The equations of motion~\footnote{For details 
see Eqs.~(5.1) and (B.6) -- (B.14) of 
Ref.~\cite{AABBS}.} are 
\beq
\frac{\delta \Gamma[\phi,G]}{\delta \phi_a(x)} = 0 \quad, \quad
\frac{\delta \Gamma[\phi,G]}{\delta G_{ab}(x,y)} = 0 \, .
\label{fieldequations}
\eeq
As we will show below, for the phenomenon of parametric resonance, 
each diagram of the infinite series
in Fig.~\ref{fig:SSBNLO} eventually contributes at the same order in 
the coupling $\lambda$. In this sense, the 2PI $1/N$-expansion
to NLO represents a minimal approach for a controlled description 
including rescattering. 
This justifies the rather involved complexity of the approximation.

\begin{figure}[t!]
\vspace*{-0.1cm}
  \begin{center}
  \epsfig{file=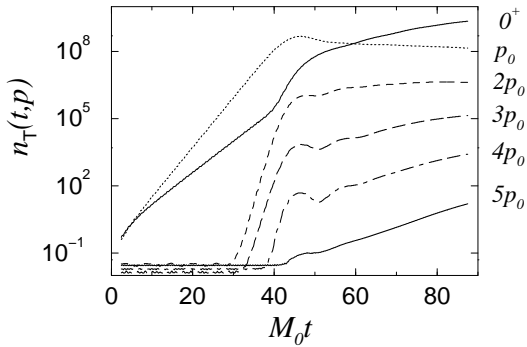,width=7.cm}
  \end{center}
\vspace*{-0.6cm}
 \caption{Effective particle number density for the 
 transverse modes as a function of time for various momenta 
 $p \le 5 p_0$. At early times, modes with $p \simeq p_0$ are  
 exponentially amplified with a rate $2 \gamma_0$. Due to 
 nonlinearities, one observes subsequently an enhanced growth
 with rate~$6\gamma_0$ for a broad momentum range.
\vspace*{-0.45cm}}
 \label{fig:number_tr}
\end{figure}

\begin{figure}[h]
\vspace*{-0.1cm}
  \begin{center}
  \epsfig{file=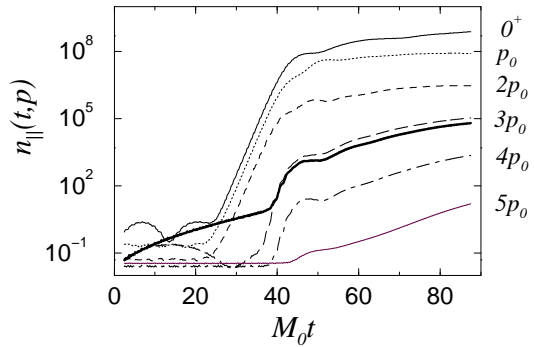,width=7.cm}
  \end{center}
\vspace*{-0.6cm}
  \caption{Same as in Fig.\ \ref{fig:number_tr}, for the longitudinal 
  modes. Nonlinear source effects trigger an exponential growth with 
  rate $4\gamma_0$ for $p \lesssim 2p_0$. 
  The thick line corresponds to a mode in the parametric resonance band, 
  and the long-dashed line for a similar one outside the band. The resonant 
  amplification is quickly dominated by source-induced particle 
  production.\vspace*{-0.4cm}}
  \label{fig:number_lg}
\end{figure}

\begin{figure}[t!]
\vspace*{-0.cm}
  \begin{center}
  \epsfig{file=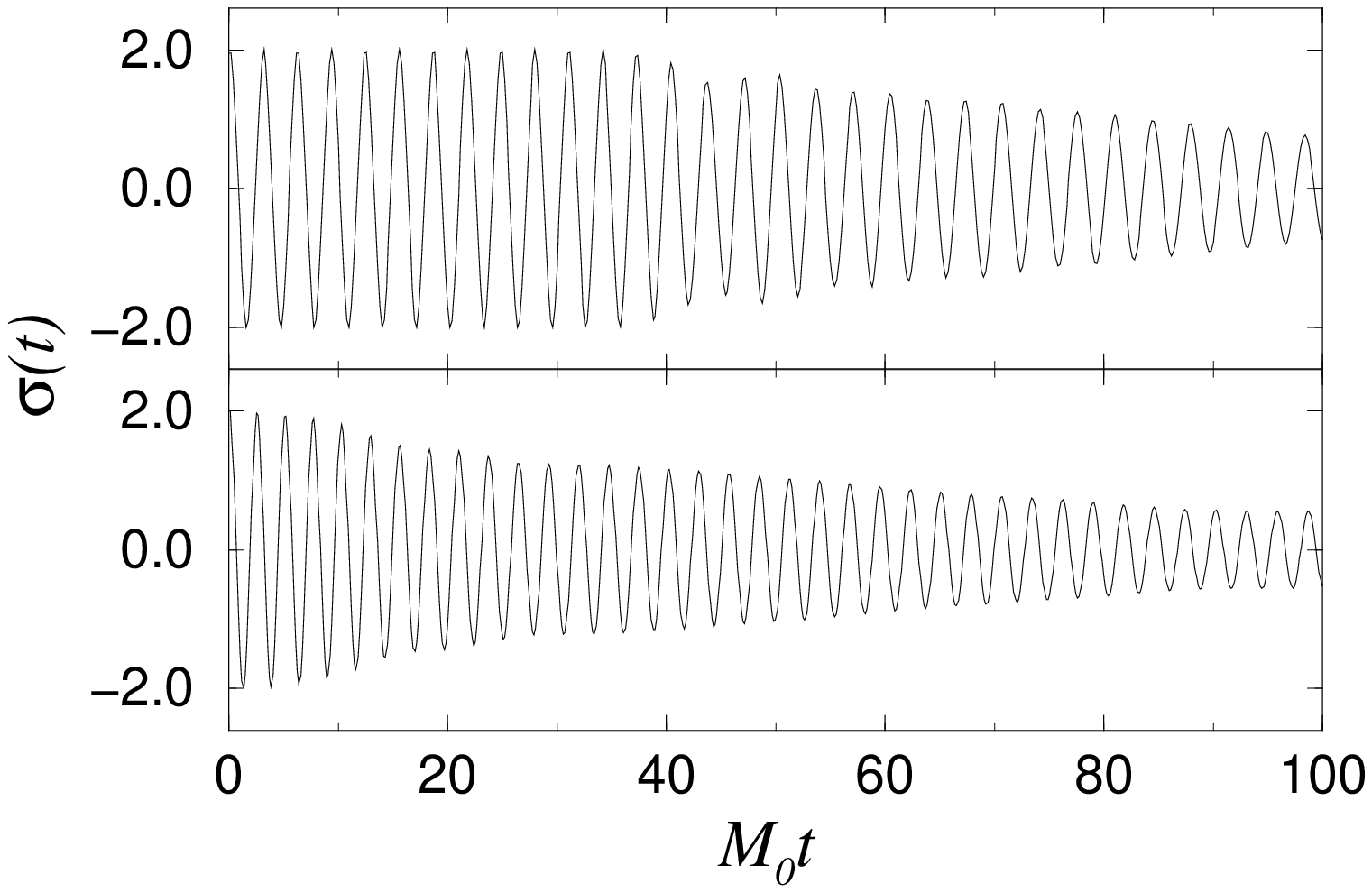,width=7.cm}
  \end{center}
\vspace*{-0.4cm}
  \caption{The rescaled field $\sigma$ as a function of time for 
  $\lambda = 10^{-6}$ (up) and $\lambda = 10$ (below).\vspace*{-0.4cm}}
  \label{fig:field}
\end{figure}

{\em Overview.} 
Parametric amplification of quantum fluctuations can be best  
studied in a weakly coupled system that is initially 
in a pure quantum state, characterized by a large 
``classical'' field amplitude    
\mbox{$\phi_a(t)\!=\!\sigma(t) M_0 \sqrt{6 N/\lambda}\, \delta_{a1}$}, and 
small quantum fluctuations, corresponding to vanishing particle numbers at 
initial time. Here $M_0$ sets our unit of mass  
and the rescaled field $\sigma(t=0)\equiv \sigma_0$ is of order unity. 
\begin{figure}[t!]
\vspace*{-0.2cm}
  \begin{center}
  \epsfig{file=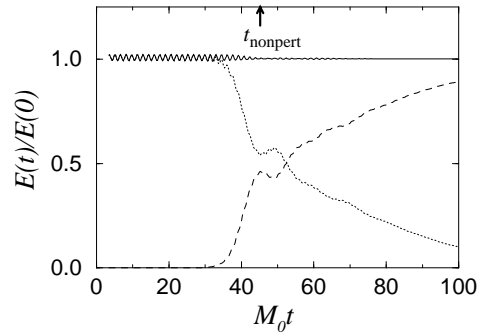,width=4.5cm,angle=-90}
  \end{center}
\vspace*{-0.6cm}
  \caption{Total energy (solid line) and 
  classical-field energy (dotted line) as a function of time for
  $\lambda =10^{-6}$. The dashed line represents the fluctuation part,
  showing a transition from a classical-field to a
  fluctuation dominated regime.\vspace*{-0.4cm}}
  \label{fig:energy}
\end{figure}  
We first present results from a full 
numerical~\footnote{Typical volumes $(N_s a_s)^3$ of $N_s = 36$--$48$ 
with $a_s = 0.4$--$0.3$ lead to results which are rather 
insensitive to finite-size and cutoff 
effects. Displayed numerical results are for $N=4$. For details about the 
numerical method see~\cite{quantum}.} integration of the time 
evolution at NLO. 
In Fig.~\ref{fig:energy} we show the classical-field and the 
fluctuation contributions to the (conserved) total
energy as functions of time. 
The former dominates at early times and one observes that
more and more energy is converted into fluctuations as the system evolves.
A characteristic time -- denoted 
as $t_{\rm nonpert}$ in Fig.~\ref{fig:energy} --
is reached when the classical and the fluctuation part of the energy 
become of the same size. Before this time,
the coherent oscillations of the field $\phi$ lead to a resonant enhancement 
of fluctuations in a narrow range of momenta around 
a specific value $|\bp |\simeq p_0$: this is parametric 
resonance~\cite{preheat1,LOapp}. 
Nonlinear interactions between field modes then cause this amplification to 
spread to a broad range of momenta. We find that the resulting rate
of amplification exceeds the characteristic 
rate $\gamma_0$ for the resonant growth. This is illustrated in 
Figs.~\ref{fig:number_tr} and~\ref{fig:number_lg}, 
where the effective particle numbers~\cite{thermal} are displayed
for various momenta as a function of time, both for 
the transverse ($G_{\pe}$) and the longitudinal ($G_{\pa}$)
sector, with $G_{ab}= {\rm diag} 
\{G_{\pa},G_{\pe},\ldots,G_{\pe}\}$. One observes that,
in contrast to the rapid changes for particle numbers
before $t_{\rm nonpert}$, a comparably slow quasistationary 
evolution takes place at later times.
The fluctuation dominated regime for $t \gtrsim t_{\rm nonpert}$ 
is characterized
by strong nonlinearities. For instance, from Fig.\ \ref{fig:energy} one infers
for $t \simeq t_{\rm nonpert}$ that the classical field decay ``overshoots''
and is temporarily reversed by feed-back from the modes.
This can be directly seen in the evolution for the rescaled field shown in 
Fig.~\ref{fig:field}. The particle numbers of Figs.~\ref{fig:number_tr} 
and~\ref{fig:number_lg} exhibit correspondingly a reverse behavior. Very 
similar phenomena have been observed in related classical 
theories~\cite{preheat2}, though a direct comparison requires
simulations for the same model as studied here~\cite{pretherm}. 
There the full nonlinearities,  
i.e.\ including {\em all orders} in $1/N$, are taken into account
while leaving out quantum corrections. This indicates 
the capability of the 2PI $1/N$ expansion at NLO to capture the 
dominant nonlinear dynamics. 

The characteristic properties described above can be understood analytically
from the evolution equations for the Fourier modes of the 
one- and two-point functions. 
Separating real and imaginary parts with
$ G_{\pa,\pe}(t,t';\bp)=F_{\pa,\pe}(t,t';\bp)-\frac{i}{2}\, 
 \rho_{\pa,\pe}(t,t';\bp)\, {\rm sign}_{\C}(t-t')$,
the real $\rho_{\pa,\pe}$ denote the spectral and $F_{\pa,\pe}$ the 
statistical two-point functions~\cite{thermal,quantum}. We define: 
\beq
 M^2(t) = m^2 + \frac{\lambda}{6N} [3 T_{\pa}(t) + (N-1) T_{\pe}(t)] \, ,
 \label{MFmass}
\eeq
where $T_{\pa,\pe}(t) \!\!=\!\! \int^{\Lambda}\! \frac{{\rm d} \bp}{(2\pi)^3}
F_{\pa,\pe}(t,t;\bp)$
with \mbox{$\Lambda \gg p_0$}.
Initially, $F_{\pa}(0,0;\bp) = 1/2 \om_{\pa}(\bp)$,
$\partial_{t}F_{\pa}(t,0;\bp)|_{t=0} = 0$,
$\partial_{t}  \partial_{t'} F_{\pa}(t,t';\bp)|_{t=t'=0} =
\om_{\pa}(\bp)/2$, and similarly for $F_{\pe}$. The frequencies 
are $ \om_{\pa}(\bp) = [\bp^2 + M_0^2\, (1 + 3 \sigma^2_0)]^{1/2}$,
$M_0^2 \equiv M^2(0)$, and similarly for $\om_{\pe}(\bp)$ with 
$3 \sigma^2_0 \to \sigma^2_0$.  
The initial conditions for the spectral functions 
are fixed by the
equal-time commutation relations: $\rho_{\pa,\pe}(t,t';\bp)|_{t=t'} = 0$
and $\partial_{t}\rho_{\pa,\pe}(t,t';\bp)|_{t=t'} = 1$~\cite{thermal,quantum}. 
  
{\em (I) Early-time linear (Lam{\'e}) regime: parametric resonance.}
At early times the $\sigma$--field evolution equation receives the
dominant ($\calO (\lambda^0)$) contributions from the classical action 
$S$. The classical field dynamics is characterized by rapid 
oscillations with constant amplitude and period $2\pi/\omega_0$. 
The evolution equations for the two-point functions 
at $\calO (\lambda^0)$ correspond to
free-field equations with the addition of a
time dependent mass term $3 \sigma^2(t)$ for the longitudinal
and $\sigma^2(t)$ for the transverse modes. 
The dynamics in this linear regime has been extensively studied in the 
literature and is known to be described by the solution of a
Lam{\'e} equation \cite{preheat1,LOapp}: Parametric resonance manifests
itself by an exponential growth of the statistical two-point functions 
describing particle production in a narrow momentum range with 
$p^2 \le \frac{\sigma_0^2}{2}$. Averaging over the short-time scale 
$\sim \omega_0^{-1}$ 
one finds for the {\em transverse} modes for $t,t' \gg \gamma_0^{-1}$:
\beq
 F_{\pe} (t,t';\bp_0) \sim e^{\gamma_0 (t+t')} \, ,
\label{earlytime}
\eeq
for the maximally enhanced mode with $p_0\simeq \frac{\sigma_0}{2}$.
One finds a much smaller growth for the {\em longitudinal} modes 
(cf.~also Fig.\ \ref{fig:number_lg}). 
The analytic solution to $\calO (\lambda^0)$
agrees accurately with the NLO numerical results at early times. 

{\em (II) Source-induced amplification regime: 
enhanced particle production for longitudinal modes.\label{II}}
Because of the exponential growth of the transverse fluctuations
for $\gamma_0 t \gg 1$, the $\calO (\lambda^0)$ approximation eventually
breaks down at some time. Using the two-loop graphs of 
Fig.~\ref{fig:SSBNLO}, at $\calO (\lambda)$ the evolution equation for 
$F_\pa$ can be approximated by
\bea
 \label{Fparevol}
 \lefteqn{
 \Big(\partial_t^2 + {\bp}^2 + 3 \sigma^2(t) + M^2(t) \Big) 
 F_\pa(t,t';{\bp}) \simeq   } \\
 &2\alpha \sigma^2(t)\,{\rm T}_{\pe}(t) \,F_{\pa}(t,t';{\bp})\, 
 + \alpha \sigma(t)\sigma(t')\, \Pi^F_\pe(t,t';{\bp})  \, ,
 \nonumber
\eea
with $\Pi^F_\pe(t,t';{\bp}) 
= \int {\rm d} \bq F_\pe(t,t';{\bp}-\bq) F_\pe(t,t';\bq)$, 
and $\alpha=\frac{\lambda}{(2 \pi)^3}\frac{N-1}{6N}\frac{c^2}{\omega_0^2}$ with constant 
$c\sim 1$. Here, we have used that the momentum integrals are dominated
by the amplified transverse modes. In addition, we exploit the fact that, for
$t'' \simeq t \lesssim t_{\rm nonpert}$,
\beq
 F_{\pe}^2(t,t'';\bp_0) \gg \rho_{\pe}^2(t,t'';\bp_0)/4 \, .
\label{condclass}
\eeq
Its validity can be seen from~(\ref{earlytime}) and the corresponding behavior 
of $\rho_{\pe}(t,t'';\bp_0) \sim e^{\gamma_0 (t-t'')}$ for 
$t'' \lesssim t$. We emphasize that strictly 
neglecting in the evolution equations 
$\rho^2$-terms as compared to $F^2$-terms for
{\em all} modes and {\em all} times corresponds to the classical 
statistical field theory limit~\cite{quantum}.
This provides an explicit 
analytic criterion for the applicability of the classical methods 
employed in~\cite{preheat2}. For sufficiently early times, 
all momentum integrals are indeed dominated by the enhanced modes
$\bp \simeq \bp_0$ for which (\ref{condclass}) is valid.
A detailed analysis reveals that the
memory integrals appearing in the equations of motion are dominated by the
latest times $t'' \simeq t$: One can explicitely verify that 
effective locality, described by the replacement  
$\int_0^t {\rm d} t'' \longrightarrow  \int_{t - c/\omega_0}^t 
{\rm d} t''$ (cf.~Eq.~(\ref{Fparevol})), 
is valid for the time-averaged behavior
over the short-time scale $\sim \omega_0^{-1}$. We conclude that
classical statistical approximations can provide a good description 
in this regime. 

The first term on the RHS of (\ref{Fparevol}) is a NLO
contribution to the mass, whereas the second term represents a
source-term, resulting from annihilations of amplified transverse modes 
as well as stimulated emission processes. To evaluate the momentum integrals,
we use a saddle point approximation around $p\simeq p_0$, valid for 
$t,t'\gg \gamma_0^{-1}$, with $F_\pe(t,t',\bp) \simeq F_\pe(t,t',\bp_0) 
\exp [-|\gamma_0^{\prime\prime}|(t+t')(p-p_0)^2/2]$:
\bea
T_{\pe}(t) &\simeq& \frac{p_0^2\,F_{\pe} (t,t;\bp_0)}{2 
(\pi^3 |\gamma_0^{\prime\prime}| t)^{1/2}}\, , \\
\Pi^F_\pe(t,t';0) &\simeq &
\frac{p_0^2\,F_{\pe}^2(t,t';\bp_0) }{4 (\pi^3 |\gamma_0^{\prime\prime}| 
(t+t'))^{1/2}}\, . \label{sourcetermlabel}
\eea
Here we wrote the source term for its maximum at $\bp = 0$. More precisely, 
it affects all modes with $\bp \lesssim 2 \bp_0$.
An important characteristic time is reached when the mass corrections become
comparable to the classical mass term. This can be seen to happen at 
$t_{\rm nonpert} \simeq (\ln \lambda^{-1})/(2\gamma_0)$, when 
${\rm T}_{\pe} \simeq \calO(\lambda^{-1})$ (see also~\cite{LOapp}). 
Note that at this time,
both the LO and the NLO mass terms are of the {\em same} order in $\lambda$,
but with opposite sign. We point out that the source term 
in (\ref{Fparevol}) becomes important at the {\em earlier} time 
\beq
t_{\rm source} \simeq t_{\rm nonpert}/2 \, ,
\eeq
when $F_{\pe}(t,t';\bp_0) \simeq \calO (N^0 \lambda^{-1/2})$.
For $t \gtrsim t_{\rm source}$, the exponentially growing source drives
the dynamics of longitudinal modes and one finds a strong amplification
for $p \lesssim 2p_0$: $F_{\pa}(t,t';\bp) \sim 
\exp [2\gamma_0 (t+t')]$. This agrees precisely with the numerical results 
shown in Fig.\ \ref{fig:number_lg}.

{\em (III) Collective amplification regime: 
explosive particle production in a broad momentum range.}
A similar analysis can be made for the transverse fluctuations.
The approximate evolution equation for $F_{\pe}$ has a similar 
structure as (\ref{Fparevol}). Beyond the $\calO(\lambda^0)$ (Lam\'e) 
description, it receives contributions from the feed-back of the 
longitudinal modes at $\calO (\lambda)$ and from the amplified 
transverses mode at $\calO (\lambda^2)$. The corresponding mass corrections 
remain small until $t_{\rm nonpert}$, whereas the source term is 
parametrically of the form $\sim \lambda^2 F_\pe^3/N$. This leads to 
the characteristic time 
\beq
t_{\rm collect} \simeq 2\, t_{\rm nonpert}/3 
+ (\ln N)/(6 \gamma_0) \, 
\label{tfield}
\eeq
at which $F_{\pe}(t,t';\bp_0) \simeq \calO (N^{1/3}\lambda^{-2/3})$. 
Correspondingly, for $t_{\rm collect } \lesssim t \lesssim t_{\rm nonpert}$
one finds a large particle production rate $\sim\! 6 \gamma_0$ 
for a wide range of momenta, in agreement with the full NLO results
in Fig.\ \ref{fig:number_tr}. In this time range
the longitudinal modes exhibit an enhanced amplification as well 
(cf.~Fig.\ \ref{fig:number_lg}). The abundant particle production 
is accompanied by an exponential decrease of the classical-field energy.   
We emphasize that the collective amplification regime is absent 
in the LO large-$N$ approximation. Consequently, even for the transverse 
sector the latter does not give an accurate description at intermediate
times if 
$t_{\rm collect }\le t_{\rm nonpert}$, that is for $N \lesssim \lambda^{-1}$. 

{\em (IV) Nonperturbative regime: 
quasistationary evolution.} At $t \simeq t_{\rm nonpert}$ one finds
$F_{\pe}(t,t';\bp_0) \simeq \calO (N^0\, \lambda^{-1})$. 
As a consequence, there are leading contributions ($\sim \lambda^0$) to the 
dynamics coming from all loop orders (cf.~Fig.~\ref{fig:SSBNLO}).  
In particular, the evolution equations are no
longer ``local'' in the sense described under {\em (II)} and memory
effects become important. In contrast to 
the rapid resonant dynamics before $t_{\rm nonpert}$, a comparably 
slow, quasistationary evolution driven by direct scattering sets in.
We emphasize that the collective amplification regime
triggered a rapid approach to monotonously decreasing 
particle number distributions as functions of momentum, which 
become quasistationary afterwards (cf.~Figs.~\ref{fig:number_tr},
\ref{fig:number_lg}). The approach 
to true thermal equilibrium is exceedingly slow for the employed
range of couplings $\lambda = 10^{-6} - 10$, for which a non-negligible
parametric resonance regime can be observed. For phenomenological
applications it is therefore crucial that in the quasistationary, 
pre-thermal regime in the spirit of Refs.~\cite{pretherm,dcc} 
a number of aspects do not differ much from the 
late-time thermal regime~\cite{BSW}.

We thank R.\ Baier, C.\ Wetterich and T.\ Prokopec for fruitful discussions.
Numerical computations were done on the PC cluster HELICS of the 
Interdisciplinary Center for Scientific Computing (IWR), 
Heidelberg University.

\end{document}